# SIMULATING THE SOCIAL INFLUENCE IN TRANSPORT MODE CHOICES

Kathleen Salazar-Serna

Facultad de Ingeniería y Ciencias
Pontificia Universidad Javeriana
Calle 18 # 118 - 250 Avenida Cañasgordas
Cali, Valle del Cauca 760031, COLOMBIA

Lynnette Hui Xian Ng

Software and Societal Systems
Carnegie Mellon University
4665 Forbes Ave
Pittsburgh, PA 15213, USA

Lorena Cadavid
Carlos Jaime Franco

Department of computer and decision sciences
Universidad Nacional de Colombia
Av. 80 #65 - 223
Medellín, Antioquia, 050034, COLOMBIA

Kathleen Carley

Software and Societal Systems
Carnegie Mellon University
4665 Forbes Ave
Pittsburgh, PA 15213, USA

**ABSTRACT**

Agent-based simulations have been used in modeling transportation systems for traffic management and passenger flows. In this work, we hope to shed light on the complex factors that influence transportation mode decisions within developing countries, using Colombia as a case study. We model an ecosystem of human agents that decide at each time step on the mode of transportation they would take to work. Their decision is based on a combination of their personal satisfaction with the journey they had just taken, which is evaluated across a personal vector of needs, the information they crowdsource from their prevailing social network, and their personal uncertainty about the experience of trying a new transport solution. We simulate different network structures to analyze the social influence for different decision-makers. We find that in low/medium connected groups inquisitive people actively change modes cyclically over the years while imitators cluster rapidly and change less frequently.

## 1 INTRODUCTION

Agent-based simulation has become a popular technique that allows for the natural decomposition of an ecosystem into individual agents that interact with each other and the environment to achieve their goals while producing emerging behaviors (Chen and Cheng, 2010). This makes agent-based models useful for representing complex systems such as transportation and analyzing the travel behavior of individuals that make decisions influenced by the status of the system and the interactions with similar travelers (Kagho et al., 2020). Transportation systems have been modeled around the world for different purposes: traffic management, passenger flows, multimodal optimization, emission control, or road security (Wise et al., 2017). Most studies have been performed in developed countries, where the main interest lies in cars, bicycles, and public transportation, ignoring motorcycles (Bakker, 2019; Cadavid and Salazar, 2020).

In the developing world, motorcycles are more likely to be the primary mode of transportation as compared to cars. In countries such as Colombia, India, and Indonesia, it is not uncommon for multiple passengers and goods to ride on motorcycles through congested streets. They are preferred because of their low cost, versatility, and short travel times (Eccarius & Lu, 2020; Tanabe & Asahi, 2018). This popularity has alarmingly increased the number of these vehicles in circulation, likewise causing issues related to traffic congestion, road accidents, and pollution (Suatmadi et al., 2019). In contrast, motorbikes represent around 5% of the vehicle fleet in high-income countries such as the United States, Canada, and Australia, where they are generally used for leisure, tourism, and sports (Broughton and Walker, 2009).

Socioeconomic and cultural factors influence the decision process for urban travelers. Hence, results from studies carried out in developed countries are not necessarily applicable to other territories. Therefore, devising more effective transport policies - in particular those oriented toward motorcycles as the most relevant transportation means in developing countries, requires more research to understand how people make decisions about their transport mode and how social interactions influence those decisions. This research explores how the interactions of urban travelers influence the mode choice of transportation in developing countries by using an agent-based simulation that represents transport dynamics and includes motorcycles as a means of transportation. Colombia is used as a case study in this work.

Colombia is a country located in South America. Its 51 million people fuse cultures from its Spanish colonial masters, European and Middle Eastern immigrants, as well as enslaved Africans. With a GDP per capita of $6,104 and an unemployment rate of 13.9% reported by the World Bank for 2021, Colombia is classified as a developing country by the International Monetary Fund.

In Colombia, the number of motorcycles increased by 697% between 2002 and 2021 and have become the most used vehicle in the country. They currently account for 62% of the total fleet, which means that there are more than 10 million motorcycles in circulation- which represents one motorcycle for every five people (ANDEMOS, 2023). Motorbikes are mainly used by low and middle-income people to commute from home to work (DANE, 2018).

We model an ecosystem of agents that decide at each time step on the mode of transportation they would take to work. After their journey, these agents evaluate their satisfaction and depending on their decision-making strategy, choose the transport mode individually, or seek satisfaction information from peers within their social network; by using a combination of personal and peer satisfaction, the agents determine the preferred mode of transport for the next time step. This paper aims to explore the travel behavior for agents that make decisions supported on their network by analyzing the implications of using different social network structures in a transport mode shift simulation model. Experiments were carried out across different settings of social networks, simulating the different types of friendship tie formation to observe how decisions on transportation modes change with different social networks due to the social dynamics of the country. Through agent centrality metrics and network graphs, we analyze the influence and power agents have among their interactions, and compare these factors with the rate of change. We provide a comparative analysis for two different structures: preferential attachment and random networks with different degree of connectivity (density). We observe that frequency and velocity of mode shift varies depending on the tie formation and total degree centrality of travelers according to the strategy used to make decisions. This suggest that interventions to change behavior might be targeted according to the specific groups and the social dynamics of the country.

## 2 RELATED WORK

### 2.1 Agent-Based Simulation in Transportation

Agent-based simulation systems have been used in the study of transportation issues by research groups from Germany to Switzerland. These applications range from modeling driver-vehicle elements as agents to simulate and assess decision making behaviors under different sets of traffic conditions, such as congestion, lane changes, traffic light coordination, and so forth (Chen and Cheng, 2010). For instance, Doniec et. al. (2008) constructed a multi-agent behavior model that simulates vehicle drivers coordinating with others when approaching road intersections, providing anticipatory abilities for reasoning and avoiding traffic gridlocks.

Studies found in the literature indicate that interaction among travelers influences transport modal shifts, leading to the formation of sub-populations of commuters (Faboya et al., 2020). Approaches such as The Consumat (Jager & Janssen, 2012) consider social engagements between similar other travelers to allow gathering information on their opinions that may clear up uncertainties in the decision-making process. The Consumat comes from the viewpoint of agents' social and personal needs and their satisfaction and uncertainty tolerance, which thus affect the behavioral options they employ: repetition, imitation, inquiring, and deliberating.

The Modal Shift (MOSH) framework (Faboya et al., 2020) applies social interaction in an agent-based model using a network structure to analyze the adaptive travel behavior in choosing transportation modes to commute to and from a university. Their agent decision module is influenced by individual satisfaction needs as well as perceptions from the environment, simulating the unplanned behavior of travelers. The results show that comfort has a large intervention effect on transport modes, and social interactions have an effect on the adoption of travel modes.

## 2.2    Social Networks Influencing the Decision Making

Agent-based models are increasingly used to study complex human-environment interactions in a variety of areas relating to transportation, epidemiology, and information diffusion. These models have increasingly been integrated with the human social networks that inform agent decisions in real-life, for example in nuclear explosion simulations (Burger et. al., 2017).

Social networks allow for interpersonal influence on decision-making, and their structures have been largely analyzed to understand the phenomenon of how social influence could end up in coordinated activities and consensus in complex communities (Friedkin, 1998). How this influence occurs can be affected by the position of the actors and the cultural dynamics that led to the formation of the structure. Social structure is the distribution of interaction probabilities, while culture is the distribution of facts.

Changes in the culture result in changes in the interaction probabilities, which finally change the distribution of those probabilities (the social structure) (Carley, 1991). Hofstede's law describes the effects of a society's culture on the values of its members, which are thus reflected in their decision-making processes (Hofstede, 1984). This law has six dimensions, representing the parameters of society. Two of these dimensions are: individualism and uncertainty. The individualism factor reflects the degree to which societies are integrated and dependent on groups. An individual society places greater importance on personal goals, while agents in a collective society are very dependent on each other, especially for decision making. Colombia is known for its collectivist culture, as it has a very tightly-knit society, similar to other developing countries such as Vietnam or Indonesia. Hofstede's Uncertainty Avoidance Index reflects a dimension to which a culture embraces future uncertainty. Societies like China and Japan are known to have high levels of uncertainty avoidance, using structured learning and strict rules to reduce uncertainty, while the United States is an example of low uncertainty avoidance, which celebrates open-ended learning situations (Hofstede et al., 2010).

We include the idea of social networks and influence used in approaches like the Consumat to explore the impact of network structures on the decision-making process. With these, we analyze how different topologies and density parameters influence the cognitive process for the inquisitive and the repetitive agents, considering that cultural dimensions in developed and developing countries vary and influence the way people interact.

# 3 METHODOLOGY

In this section, we detail the methodology of analyzing transportation decisions influenced by the social network. We first illustrate the overview of the developed model, which is based on the Consumat approach (Jagger and Janssen, 2012) and the MOSH framework (Faboya et al., 2020). First, we explain the process that commuters follow to choose a transport mode, providing a high-level view of the model. Then, we describe the social structure implemented to connect travelers and allow their interaction. Finally, we explain the validation process and present the parameters used to run experiments.

The simulation is implemented with the NetLogo software, version 6.3.0 (Wilensky, 1999). Time is measured in ticks as per the software's convention. The unit of modeling is the agent, which is a human individual commuting between home and work.

## 3.1 Overview of the Model

Figure 1 illustrates the conceptual model of the system. At each time step, one of the three transport modes—motorcycle, car, or bus—is selected, and the agents make their journey using the selected transport mode to commute from home to work. Then, they evaluate their satisfaction with the journey based on a list of factors that reflects how a person describes his contentment with the travel experience. At the same time, the agents check the uncertainty about the satisfaction that will be obtained with their current transport mode. This is a combination of their own experience of using it and the experience of contacts in his social network, which reflects how a person can be influenced by the experiences of his immediate peers. We represent these factors with the scaled cultural dimensions of Hofstede et al., (2010): personal uncertainty avoidance and collectivism. These dimensions score Colombians as highly collectivistic people, for whom belonging to an in-group and aligning with that group's opinion are very important (Hofstede, 2023).

After the agent combines the personal information with that of his peers, he evaluates a mental model that determines the strategy to apply for the decision-making process and, thus, the transport mode choice for the subsequent tick. The flow is repeated for each agent at each decision step until the end of the analysis period. For this study, we are analyzing a time period equivalent to 10 years, that means that the agent changes a total of 10 times, one per year. We simulate a typical daily peak hour per year, and then people make shift decisions each year based on their experiences on the trip.

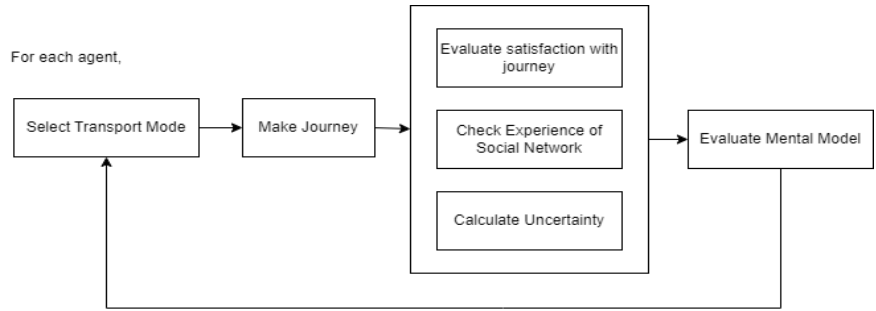

Figure 1: Overview of the conceptual model used to evaluate travel and make mode decisions.

The initial setup for transport mode is a probabilistic allocation of the three available transport modes: motorcycle, car, and public transit. This allocation depends on the initial attributes of agents: gender, age, and socioeconomic status. Our simulation was parameterized with the average distribution of users by transport mode in Cali, Colombia. Subsequently, the transport mode is dependent on the evaluation of the mental model set out in our simulation model.

Agents are initialized with a home and work location, all agents start the journey to their assigned office at the same time, representing the peak hour. After an agent completes his journey, he evaluates his satisfaction with the home-work commute. The overall satisfaction is calculated as indicated in equation (1), where $V$ is the standardized value for each *Need i* and $W$ is the weight for that Need; $n$ represents the number of Needs considered in the equation. Based on previous studies (Faboya et al., 2020; Wangsness et al., 2020), we consider the Needs as: the cost of acquisition, operating cost, road safety, personal security during travel, travel comfort, commuting time, and pollution generated. Values for these Needs are affected by the status of the system as a result of individual and aggregate decisions of the agents. Weights can be parameterized according to the cultural characteristics in the society analyzed. After agents calculate their satisfaction, we compare it to a threshold that represents the level of overall Needs satisfaction a commuter wishes to experience.

$$Overall\ satisfaction = \sum_{i=1}^{n} V_{Need\ i} * W_{Need\ i} \qquad (1)$$

Decisions are usually made under uncertainties and agents interact with similar others to obtain information that might help to diminish their uncertainty (Janssen, 2003). Travelers calculate their uncertainty level and compare it to the uncertainty threshold- this is the agent's level of tolerance towards the uncertainty. Thresholds can take values from 0 to 1. In this context, uncertainty is a mental state experienced due to a lack of certainty about the satisfaction that will be obtained with the transport mode on future trips. The more experience accumulated (individual and collective), the less uncertainty there is.

We use the Consumat to classify agents into categories depending on the strategy used to decide about the transport mode. When the agent's uncertainty is below the threshold, the reasoning process is made individually, and depending on whether the satisfaction is high or low, they repeat the selection from the previous period or deliberate about the old and all the new possibilities in the system. In contrast, high levels of uncertainty imply engaging with people in their social network using one of the two behavioral strategies: imitation or inquiry.

For agents with high satisfaction, the social comparison will result in imitation, where they adopt, without a rational process, the most popular mode among their contacts (the highest proportion of $n$ neighbors using the transport $j$); that means, they take their neighbors' behavior as a benchmark to alleviate their uncertainty. High uncertainty and low satisfaction values means agents inquire in their network and compare in a rational process, their own satisfaction to the satisfaction expected if using the others' transport modes. The model recalculates satisfaction for each agent with their own individual parameters but using the parameters of the $j$ alternatives used by all of their neighbors. If any of those expected satisfactions is greater than the current satisfaction, then the agent will choose that alternative. In our study, we are interested in analyzing how the structure and density of the social network influence the mode shift for the inquirers and imitators.

### 3.2 Social Network

The human decision-making process is typically influenced by their social connections, or social network ties. Social connections have been observed to sway opinions on different kinds of issues. These changes in opinions are of concern as they can manifest in real-life decisions: the tendency to stay with or withdraw from an infrastructure project (Li et. al.), formation of housing and social clusters (Burger et. al., 2017), the refusal to take life-saving vaccinations (Ng and Carley, 2022) or evaluating modal shifts (Zhang et al., 2022).

We factor in observations of transport modes by an agent's social network as an influencing aspect of their choice of transport mode in the next tick cycle. A social network is created using the Netlogo extension to connect travelers. We tested two different types of social network topology: preferential attachment and random network topology. Previous work indicates that large social networks follow a scale-free power-law distribution (Janssen, 2003) which has been applied to analyze market dynamics and could be also extended to transport services. The preferential attachment topology is a common model for real-world networks that follow the power-law, in which agents connect to m other randomly chosen nodes by giving preference to those that are already well connected. The random topology assumes agents choose their partners uniformly at random throughout the network (Doerr et. al., 2012). We want to contrast the impact on transport selection dynamics when only a small number of nodes are highly connected versus a uniform degree distribution. For each type of network, we run the simulation with different levels of density, varying the minimum degree (number of links) for the preferential attachment network and the probability of connection for nodes in the random network. We compare low densities to represent societies that are more individualistic, like in developed countries, against medium and high densities that are more characteristic of developing countries.

### 3.3 Validation

The model was validated using the "validation in parts" technique, which means that inputs, processes, and outputs are validated separately (Carley, 2017). Inputs should have the same distributional properties as the real world. In this case, Census Data for 2018 from Cali was used to parametrize the initial distribution of the agent's properties: age, gender, socioeconomic situation, and the probability of having a motorcycle given those characteristics. Internal processes in the model need to somewhat resemble processes seen in the real world, qualitatively matching elements in the model with an analog element in the real world. To do so, first, a conceptual model was defined based on the literature of transport modeling and human behavior theories. Then the model was implemented using the Netlogo language, and code procedures were incrementally validated in each of the modules (setup and go procedures). By using extreme values of parameters as proposed by Sargent (2000), the consistency of code was validated. Experts in agent-based simulation and transportation studies were consulted to check the assumptions made in the model as a conceptual validation. Regarding outputs, the mean prediction of the model should agree with the general pattern of behavior in the real system or historic examples. For the pattern modeling validation as suggested by Railsback and Grimm (2019), a tailored scenario that represented the proportion of people using cars and motorcycles in 2016 showed that emerging macro level outputs are comparable to real-world patterns in subsequent years.

### 3.4 Experiments

We performed a series of virtual simulation experiments with a synthetic population of 100 agents. We consider social network type and density as independent variables to evaluate the impact of the different structures. Density was set at 1 (low), 3 (medium), and 5 (high) for the minimum degree of preferential attachment networks. Similarly, for the random networks, we used values of 0.05, 0.08, and 0.15. By setting the satisfaction threshold at 1 and the uncertainty threshold at 0, all agents become inquisitive. A setup of 0 for both thresholds makes all people imitators. The Weights of Needs can take values between 0 and 1, which in these experiments are considered all equal and set at one. Empirical data were collected for Cali, Colombia, as a reference to inform the model and parameterize it with a city in a developing country. A complementary description of the parameters and values used for the need scores to calculate journey satisfaction can be found in the supplementary material.

## 4 RESULTS

This section presents the results of using different network structures to connect inquisitive and imitative people that are making decisions about transport modes.

## 4.1 Inquisitive Agents

For each type of synthetic social network, we generated them with differing degrees, which simulates the density of connected people Figure 2 shows average mode shifts for preferential attachment and random networks for some periods; graphs for all of the time steps can be found in supplementary material. We found that in preferential attachment networks, during the first steps, there is a higher number of agents changing their transport mode (red nodes), either they are central or peripheral nodes. With a low minimum degree, nodes in the periphery have fewer total changes than those with the highest total centrality. In populations with a higher degree of connectivity, the majority of changes occur in peripheral nodes.

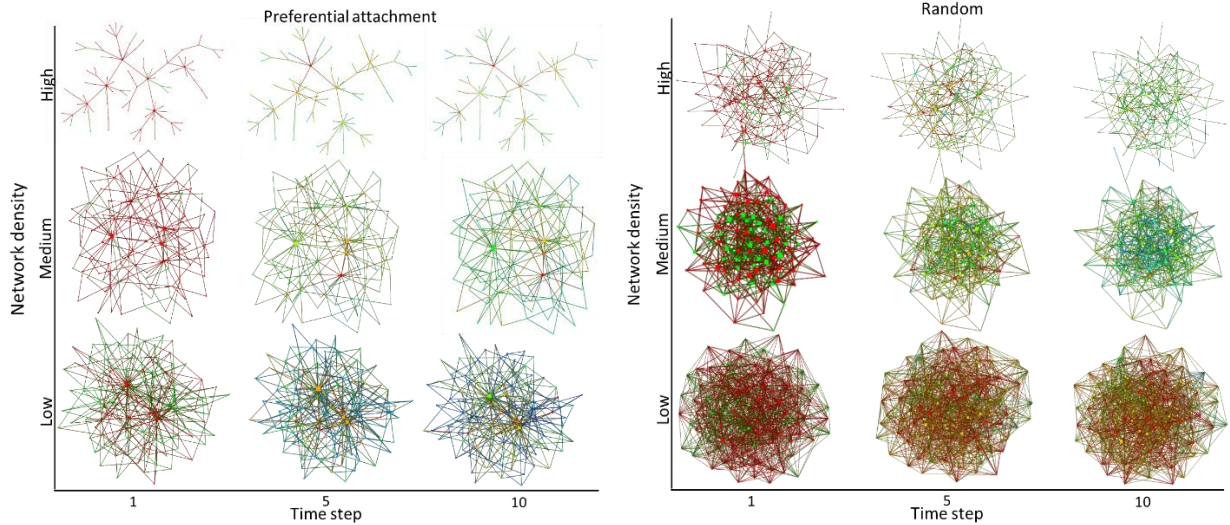

Figure 2. Illustration of social network diagrams over time steps. Agents are colored by the number of accumulated changes at tick t and sized by total degree centrality. Color scale varies from blue (low) to red (high).

In random networks, positionality is not as important as it is in preferential attachment, and as it can be seen in Figure 2, over the 10-time steps, changes occur either in central or peripheral nodes. The denser the network is, the more changes the agents make.

In both random and preferential attachment networks, inquisitive travelers actively change their transport mode every period due to their constant dissatisfaction. Yet, the rate of change is higher for more dense networks. Accumulated changes are on average higher for preferential attachment networks than random networks.

For networks with a higher degree, most changes occur during the first periods of simulation; people migrate to one of the transportation modes in greater proportion, and distribution tends to converge to one mode in the later periods. In contrast, in low-degree networks, the distribution of transport modes does not change drastically over the 10 years.

Figure 3 depicts the average distribution of the fleet in the city if all of the agents are inquisitive people. By using the setup of Cali City, we found that whether the network is random or preferential attachment there is an oscillatory behavior where people move mainly between motorcycle and bus, according to the satisfaction values obtained at each period. This behavior is explained by the constant dissatisfaction and high uncertainty as result of the comparison against the thresholds set as zero for uncertainty acceptance and one for satisfaction in this experiment. The observed pattern reflects the intention of change due to the discontentment with the current mode. Nevertheless, the socioeconomic conditions in developing countries could make it difficult to actually shift.

For highly connected people, after several periods, all agents eventually converge on this transport mode. People with medium and low degree of connection present fewer total changes despite continuing to change without convergence to a single mode. Mainly in low degree networks, we can observe a segmentation in the transport modes.

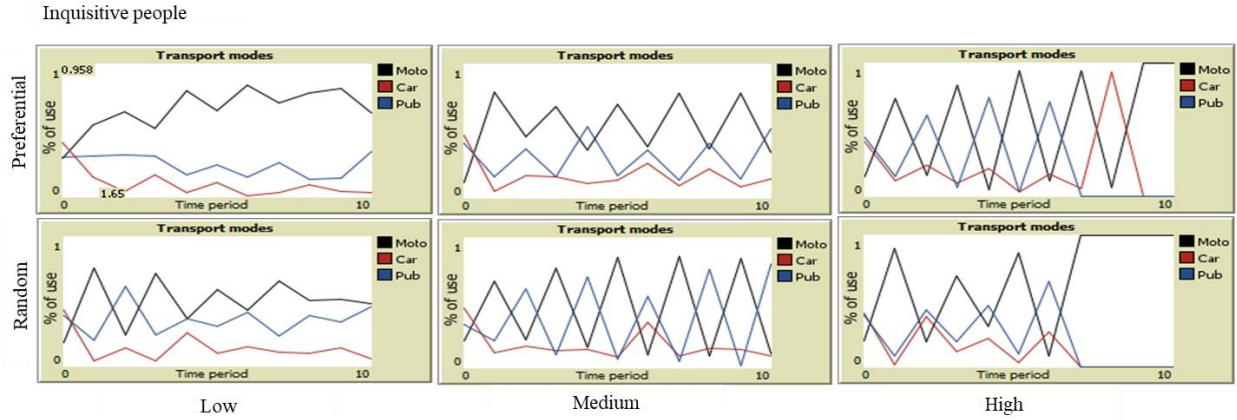

Figure 3. Average distribution of transport modes across time for inquirers with different social structures

## 4.2 Imitative Agents

Similar in manner to inquisitive agents, we ran simulations and analyzed network graphs for imitative agents. In this case, agents in low-density networks have undergone the most changes. It is due to the fact that agents with a high connection degree early imitate the behaviors that the majority of people in their network exhibit; therefore, everybody is rapidly using the same transport mode. This convergence on one transport mode occurs faster in random networks than in preferential attachment structures.

For a high connection degree, convergence is achieved in the first period, while for a medium degree, there are no more changes after the second or the third period. To a low degree, agents keep changing transport modes, and convergence does not occur within the simulated time period for preferential attachment networks, while it occurs at random networks.

In Figure 4, it is possible to observe that people in random networks copy their peers' behavior faster than those in preferential attachment structures. Although this pattern is evident at all of the connection levels, it occurs earlier with higher densities, and once all of the contacts in the network end up using the same mode, people stop changing. In the city analyzed, most people are users of public transportation services.

The initial setup in the simulation follows that distribution. Therefore, for highly connected people, it is common to have more contacts in their network using public transportation, which makes it more likely to form homogeneous groups during the first periods of simulation. On the other hand, people in more sparse networks maintain a higher level of heterogeneity, and people continue changing until the end of the time period.

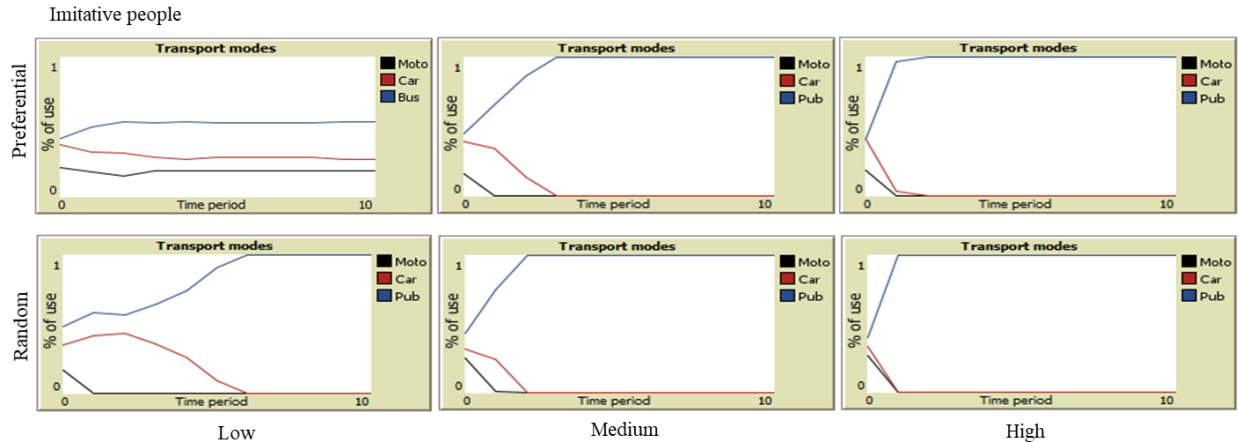

Figure 4. Average distribution of transport modes across time for imitators with different social structures

## 4.3 Discussion

Motorcycles have become the most popular mode of transportation in Colombia. The large number of these vehicles on the road has negative impacts on environmental pollution, traffic congestion, and road accidents. While there have been studies on transportation decisions around the world, most are performed for developed countries and use cars and larger transportation devices as the primary agent mode, ignoring the two-wheeled motorcycle as a full transportation mode that developing countries use. This work aims to contribute to that lack of knowledge as a starting point to analyze transport policies for the Global South.

Our simulations provide a glimpse of this phenomenon. The denser the network, the more changes the agents make, showcasing that agent transport mode decisions are truly influenced by their social network. However, after a period of time, the number of agents taking each transport mode converges regardless of the degree of connectivity.

Motorcycles oscillates with the greater proportion between 20% and 80% when people behave as inquirer, while public transportation remains as the main mode for imitators. These findings are consistent with Janssen and Jager (2003), which show that people in more cohesive groups acquire the same goods/services more frequently. Denser social networks foster homogeneity in consumption, while sparse networks present more segmentation and heterogeneity, as concluded by McPherson & Smith-Lovin (2001) in their research about homophily. Furthermore, such convergence reflects the cultural dynamics and factors of Colombian society: a smaller, light-weight mode of transportation is preferred over public transportation and cars. That predominance is a result of the satisfaction evaluation where aspects such as low costs, high insecurity rate in the public transportation and relative higher speed to cars, might have positive effects on the election of the motorcycle, despite a high accident rate. Although this paper focuses on the social structure impact, an extension of this work (Salazar-Serna et al., 2023) which is currently under review, evaluates the impact of the above-mentioned factors on the decision-making process.

A current limitation of the model is that people make the change decision every year and that might be less realistic, especially for developing contexts due to the socioeconomic conditions. This issue is further considered in our work in progress. Much of this paper makes use of simulated information. While such information is designed to represent real world values, they are still best guesses and fall short of the full spectrum of factors that humans use in their decision-making process. Future work that is in the pipeline involves a mixed-methods study that collects data from survey participants to better calibrate some parameters in the system such as the needs weights. Also, further analysis is required running simulations using different thresholds to include all the decision-maker groups simultaneously to explore dynamics of the whole system.

# 5 CONCLUSION

In this work, we constructed a model for simulating an ecosystem of transportation decisions in developing countries. The model considers motorcycles as one of the alternatives for primary transport modes since they are the preferred commuting option in the developing world and little research has been done on them. Public policies are needed to compensate for the negative impacts of overflowed motorcycle traffic on urban mobility, and our model can be used as a testbed to analyze future policies.

We created a social-influence based model where agents balance their satisfaction and uncertainty about their transportation modes with their observations of their neighbors. The simulation was parameterized using the information of Cali as a case study, which is a city located in Colombia.

The results presented in this paper indicate that not only the satisfaction of the transport needs but also the type of social network with which they share information to alleviate high uncertainty and the network density influence the modal shift. Unsatisfied people in densely connected societies, such as those in developing countries, change their modes more frequently, influenced by the constant comparison with their peers. The homophily phenomenon makes people adopt one of the modes in greater proportion. This convergence happens in the short run for people that imitate others' behavior, while it takes a longer time for inquisitive people to be coordinated. Convergence by social influence to one mode or another depends on the factors associated with the satisfaction of needs. The simulation results show that for a city in a developing country where the parameters were setup with low acquisition/operating costs are and high insecurity for public transportation, the selection favors the motorcycle.

This suggests that changes in mode adoption behavior can be driven by policymakers targeting different strategies according to the specific groups of decision-makers and the social dynamics of the country.

We hope this paper sheds light on factors that influence the decision-making process for transportation mode usage within a developing society and opens up conversations for academic researchers and policymakers to analyze the developing world.

## ACKNOWLEDGMENTS

We thank CASOS at Carnegie Mellon University and the CBIE at Arizona State University for providing support to Kathleen Salazar during her visit as a PhD student at Universidad Nacional de Colombia. We would also like to extend our appreciation to the Fulbright Scholarship Program for supporting Kathleen's visits to both universities. This program played a vital role in enabling her to pursue her academic goals and make significant contributions to our research project. Views and conclusions are those of the authors and do not necessarily represent the views of official policies.

## SUPLEMMENTARY MATERIALS

Additional information can be found at: https://github.com/Kathleenss/WSC2023-SupplementaryMaterial

## REFERENCES

Asociación Nacional de Movilidad Sostenible ANDEMOS. 2023. Cifras y Estadísticas. Informe interactivo sector automotor 2023. https://www.andemos.org/informes-interactivos

Bakker, S. 2019. "Electric Two-Wheelers, Sustainable Mobility and the City". Sustainable Cities - Authenticity, Ambition and Dream, November 2018. https://doi.org/10.5772/intechopen.81460

Broughton, P., and Walker, L. 2009. "Motorcycling and Leisure, Understanding the Recreational PTW Rider". 1st ed. Boca Raton, Florida: Routledge.

Burger, A., Oz, T., Crooks, A., and Kennedy, W. G. 2017. "Generation of realistic mega-city populations and social networks for agent-based modeling". In *Proceedings of the 2017 International Conference of The Computational Social Science Society of the Americas*. October 19th-22nd, Santa Fe, NM, USA, 1-7.

Cadavid, L., and Salazar-Serna, K. 2021. "Mapping the Research Landscape for the Motorcycle Market Policies: Sustainability as a Trend—A Systematic Literature Review". *Sustainability*, 13(19), 10813.

Carley, K. 1991. "A Theory of Group Stability". *American Sociological Review*, 56(3), 331–354. https://doi.org/10.2307/2096108

Carley, K.:Validating Computational Models. CASOS technical report. 2017. http://www.casos.cs.cmu.edu/publications/papers/CMU-ISR-17-105.pdf, accesed 13th June, 2023.

Chen, B., and Cheng, H. H. 2010. "A Review of the Applications of Agent Technology in Traffic and Transportation Systems". *IEEE Transactions on Intelligent Transportation Systems 11(2): 485-497*

DANE - Departamento Nacional de Estadística. 2019. Encuesta nacional de calidad de vida (ECV) 2019. https://www.dane.gov.co/index.php/estadisticas-por-tema/salud/calidad-de-vida-ecv/encuesta-nacional-de-calidad-de-vida-ecv-2019, accessed 8 April 2023.

Davis, P.K., O'Mahony, A., Pfautz, J. and Carley, K.M. 2019. "Social-Behavioral Simulation: Key Challenges". In *Social-Behavioral Modeling for Complex Systems* (eds P.K. Davis, A. O'Mahony and J. Pfautz). https://doi.org/10.1002/9781119485001.ch32

Doerr, B., Fouz, M., and Friedrich, T. 2012. "Why rumors spread so quickly in social networks". *Communications of the ACM 55(6): 70-75*

Doniec, A., Mandiau, R., Piechowiak, S., and Espie, S. 2008. "A behavioral multi-agent model for road traffic simulation". *Engineering Applications of Artificial Intelligence 21(8): 1443-1454*

Eccarius, T., and Lu, C.-C. 2020. "Adoption intentions for micro-mobility – Insights from electric scooter sharing in Taiwan". *Transportation Research Part D: Transport and Environment*. 84: 102327.

Faboya, O. T., Ryan, B., Figueredo, G. P., and Siebers, P.-O. 2020. "Using agent-based modelling for investigating modal shift: The case of university travel". *Computers and Industrial Engineering*, 139:106077

Friedkin, N. 1998. "A Structural Theory of Social Influence (Structural Analysis in the Social Sciences)". Cambridge: Cambridge University Press. doi:10.1017/CBO9780511527524

Hofstede, G. 1984. "Culture's Consequences: International Differences in work-related values". 2nd ed. Beverly Hills, Cal: Sage.

Hofstede, G., Hofstede, G. J., and Minkov, M. 2010. "Cultures and organizations: Software of the mind". 3rd edition. New York: Mcgraw-hill

Hofstede, G. J., 2023. Hofstede Insights: Country Comparison Tool. http://word.tips.net/Pages/T000273_Numbering_Equations.html, accessed 14th June 2023.

Janssen, M. A., & Jager, W. 2003. "Simulating market dynamics: interactions between consumer psychology and social networks". *Artificial life*, 9(4), 343–356. https://doi.org/10.1162/106454603322694807

Jager, W., and Janssen, M. 2012. "An updated conceptual framework for integrated modeling of human decision making: The Consumat II". In *Workshop Complexity in the Real World@ECCS*. Sept 3rd-7th, Brussels, 1-18.

Kagho, G. O., Balac, M., and Axhausen, K. W. 2020. "Agent-Based Models in Transport Planning: Current State, Issues, and Expectations". In *The 9th International Workshop on Agent-based Mobility, Traffic and Transportation Models, Methodologies and Applications.* Apr 6th-9th, Warsaw, Poland, 726-732.

Li, W., Yuan, J., Ji, C., Wei, S., and Li, Q. 2021. "Agent-based simulation model for investigating the evolution of social risk in infrastructure projects in China: a social network perspective". *Sustainable Cities and Society 73:103112*

McPherson, Miller, Lynn Smith-Lovin, and James M. Cook. 2001. "Birds of a feather: Homophily in social networks." *Annual review of sociology* 27, 415–444. http://www.jstor.org/stable/2678628

Ng, L. H. X., and K. M. Carley. 2022. "Pro or Anti? a Social Influence Model of Online Stance Flipping". *IEEE Transactions on Network Science and Engineering: 1-18.*

Railsback, S. F., & Grimm, V. 2019. *Agent-Based and Individual-Based Modeling. A Practical Introduction* 2nd ed. Princeton University Press. http://www.railsback-grimm-abm-book.com/book-objectives-2nd-edition/

Salazar-Serna, K., Cadavid, L., Franco C.J. and K. M. Carley. 2023. "Simulating Transport Mode Choices in Developing Countries" In *Proceedings of the 16th International Conference on Social Computing, Behavioral-Cultural Modeling, & Prediction and Behavior Representation in Modeling and Simulation*, September 20th-22nd, Pittsburgh, PA, USA.

Sargent, R. G. 2000. "Verification, validation and accreditation of simulation models". In *Proceedings of the 2000 Winter Simulation Conference Proceedings* (Cat. No.00CH37165). December 10th-13th, Orlando, FL, USA, 2000, 50-59 vol.1, doi: 10.1109/WSC.2000.899697.

Suatmadi, A.Y., Creutzig, F., & Otto, I. M. 2019. "On-demand motorcycle taxis improve mobility, not sustainability". *Case Studies on Transport Policy*, 7(2), 218–229. https://doi.org/10.1016/j.cstp.2019.04.005

Tanabe, R., and Asahi, Y. 2018. "Analysis of trends of purchasers of motorcycles in Latin America". In *Proceedings of the 20th International Conference on Human Interface and the Management of Information*. July 15th-20th, Las Vegas, Nevada, USA, 136-144.

Wangsness, P., Proostb, S., Løvold K. "Vehicle choices and urban transport externalities". *Transportation Research Part D*, 86, 102384 (2020). https://doi.org/10.1016/j.trd.2020.102384

Wilensky, U. 1999. "NetLogo". http://ccl.northwestern.edu/netlogo/. Center for Connected Learning and Computer-Based Modeling, Northwestern University, Evanston, IL.

Wise, S., Crooks, A., & Batty, M. 2017. "Transportation in Agent-Based Urban Modelling BT - Agent Based Modelling of Urban Systems. In M.-R. Namazi-Rad, L. Padgham, P. Perez, K. Nagel, & A. Bazzan (Eds.), *ABMUS: International Workshop on Agent Based Modelling of Urban Systems 2016* (pp. 129–148). Springer International Publishing. https://doi.org/10.1007/978-3-319-51957-9

Zhang, Q., Liu, J., Yang, K., Liu, B., & Wang, G. 2022. "Market adoption simulation of electric vehicle based on social network model considering nudge policies". *Energy*, 259, 124984. https://doi.org/https://doi.org/10.1016/j.energy.2022.124984

## AUTHOR BIOGRAPHIES

**KATHLEEN SALAZAR-SERNA** is a PhD candidate in the Department of Computer and Decision Sciences at Universidad Nacional de Colombia Sede Medellín and an assistant professor at the School of Engineering and Sciences at Pontificia Universidad Javeriana in Cali. Her current research interest focuses on complex systems analysis, including sustainability issues and transport policy analysis. She uses agent-based modeling and network analysis to study transport dynamics. Her email address is kgsalaza@unal.edu.co, and her website is https://orcid.org/0000-0003-3824-7044

**LYNNETTE HUI XIAN NG** is a PhD student in the Software and Societal Systems in the School of Computer Science at Carnegie Mellon University. Her primary research interest is in the area of agent-based modeling within social networks. She utilizes simulations to analyze online discourses. Her email address is lynnetteng@cmu.edu, and her website is https://quarbby.github.io.

**LORENA CADAVID** is a professor in the Department of Computer and Decision Sciences at Universidad Nacional de Colombia Sede Medellín. In addition to her academic role, she is an enterprise consultant who applies her expertise to guide organizations towards data-driven decision making. Her research interest lies in policy design through modeling and simulation of social phenomena, and she leverages data analysis to support entrepreneurial decision making. Her email is dlcadavi@unal.edu.co and her website can be found at https://orcid.org/0000-0002-6025-5940

**CARLOS JAIME FRANCO** works as a full professor in the Department of Computer and Decision Sciences at Universidad Nacional de Colombia Sede Medellín. His research areas include complex systems, energy market modeling and simulation, and policy evaluation and strategy formulation. His email is cjfranco@unal.edu.co.

**KATHLEEN CARLEY** is a professor at the Institute for Software Research, Carnegie Mellon University, the Director of the Center for Computational Analysis of Social and Organizational Systems (CASOS), the Director of the Center for Informed Democracy and Social Cybersecurity (IDeaS), and the CEO of Netanomics. Her research interests include computer science and social science for addressing complex real-world issues, such as social cybersecurity, disinformation, disease contagion, disaster response, and terrorism, from a high dimensional network analytic, machine learning, and natural language processing perspective. Her email address is kathleen.carley@cs.cmu.edu, and her website is http://www.casos.cs.cmu.edu/carley.html